\documentclass[twocolumn]{aastex631}
\usepackage{xcolor}
\usepackage{threeparttable}
\definecolor{R}{RGB}{255, 0, 0}
\definecolor{B}{RGB}{0,0, 255}
\usepackage{amsmath}

\usepackage{booktabs}
\shorttitle{Optical origin of MAXI J1348-630}
\shortauthors{Fan et al.}
\graphicspath{{./}{figures/}}

\begin{document}

\title{On the optical emission in the mini-outburst of the black hole X-ray binary MAXI J1348-630}

\correspondingauthor{Bei You}
\email{youbei@whu.edu.cn}

\author[0000-0001-7350-8380]{Xiao Fan}
\author[0000-0002-8231-063X]{Bei You}
\affiliation{Department of Astronomy, School of Physics and Technology, Wuhan University, Wuhan 430072, People's Republic of China}
\author{Dizhan Du}
\affiliation{Institute for Astronomy, School of Physics, Zhejiang University, Hangzhou 310058, China}
\author[0009-0007-7292-8392]{Han He}
\author[0009-0007-6790-437X]{Shuaikang Yang}
\affiliation{Department of Astronomy, School of Physics and Technology, Wuhan University, Wuhan 430072, People's Republic of China}

\begin{abstract}
We investigate the optical emission of the black hole X-ray binary MAXI J1348–630 during its 2019 mini-outburst. Using optical data from the Las Cumbres Observatory Global Telescope and X-ray data from Insight-HXMT, we performed time delay analysis, optical-X-ray correlation analysis, and spectral energy distribution (SED) fitting. Our key findings are as follows: (1) The X-ray Comptonization flux lags behind the optical emission by $\sim 8.5$ days, a delay naturally explained by the disk instability model (DIM). (2) The optical and X-ray fluxes show a power-law correlation with a slope $\sim 0.4$, which lies between the predicted values for viscous heating and X-ray reprocessing, consistent with the DIM framework. (3) SED fitting with the irradiated disk model successfully reproduces the quasi-simultaneous optical and X-ray data, and the contribution of the jet is negligible. Our results indicate that the optical emission during the mini-outburst originates from the disk, rather than the jet or hot accretion flow, and highlight the critical role of the DIM in understanding the mini-outburst of X-ray binaries.
\end{abstract}


\section{Introduction}
%双星及其辐射简介
During an outburst of low-mass X-ray binaries (LMXBs), a compact object (e.g., black hole in this work) accretes material through Roche lobe overflow from a companion star with mass $\lesssim 1 \, M_{\odot}$, emitting large amounts of radiation from radio to the X-ray band \citep[see][for review]{Belloni2016}. LMXBs are characterized by their strong X-ray emission, consisting of a thermal component originating from the inner region of the accretion disk \citep{SSD1973,Mitsuda1984} and a nonthermal component generally arising from a hot accretion flow or a Comptonizing corona \citep{Zdziarski2004,Sunyaev1980}. Radio emissions from LMXBs are commonly attributed to synchrotron radiation originating from powerful collimated jets \citep{Hjellming1988,2006csxs.book..381Fender}. However, the origin of optical emission in LMXBs is still not fully understood, as multiple mechanisms can contribute to, including the viscous heating of the accretion disk \citep{Yang2022,You2023sience}, the irradiation of the outer disk by X-rays \citep[i.e., X-ray reprocessing;][]{Paradijs1994,Rykoff2007,diskir2008,diskir2009}, the optically thin synchrotron radiation from jets \citep{Fender2001,Markoff2001,Homan2005opt,Russell2006}, and the synchrotron radiation from the hot accretion flow \citep{Veledina2013,Poutanen2014SSRv,Kajava2016}.

DIM has been widely adopted to explain the outbursts observed in LMXBs \citep{Lasota2001,2001Dubus,Hameury2020}. In this framework, thermal-viscous instabilities are triggered at some place when the temperature reaches the hydrogen ionization temperature, causing a rapid increase in the temperature and mass accretion rate, which results in an outburst. Consequently, the temperature and surface density of the ionized and neutral regions undergo sharp variations in a small region known as the cooling (ionized to neutral) or heating (neutral to ionized) front. The propagation of heating and cooling fronts governs the evolution of an outburst, in which the inward-moving heating front initiates the outburst and the subsequent cooling front drives its decay \citep[e.g.,][]{2001Dubus}. 

%态转换
LMXBs often exhibit state transitions in X-rays, shifting between different accretion states \citep[see][for review]{Remillard2006ARAA}. These transitions are typically classified into the ``low-hard state (LHS)", where the emission is dominated by the thermal Comptonization in a corona, and the ``high-soft state (HSS)", where the thermal emission becomes prominent \citep{Homan2005state}. During an outburst, LMXBs typically transition from the LHS through an intermediate state to the HSS, and eventually return to the LHS, tracing a ``q-shaped" trajectory in the hardness–intensity diagram \citep[e.g.,][]{Homan2001}. However, this canonical pattern is not universal: approximately 30\% -- 40\% of outbursts either remain in the LHS throughout the entire period or enter the intermediate state but fail to transition to the HSS \citep[so called the ``failed outbursts",][]{Alabarta2021,Lucchini2023}, which was first noticed by \citet{Brocksopp2004}. The truncated disk model serves as the paradigm for understanding state transitions \citep{ARAADone2007,ARAAYuan2014}, and therefore understanding what governs the state transitions remains a central question in black hole accretion physics, which reflect fundamental changes in the geometries and dynamics of the accretion flow \citep[e.g.,][]{Dai2020,You2021,You2023sience}. 

%介绍minioutburst
It is not uncommon for LMXBs to exhibit one or more secondary maxima in their light curves after the main outburst, typically with fluxes 1-2 orders of magnitude lower than the main outburst. These phenomena are often referred to as mini-outbursts \citep[also known as reflares and re-brightnings][]{Zhang2019,Ozbey2022,2023Saikia}. The majority of mini-outbursts do not reach the HSS, while a subset could experience state transitions \citep[e.g.,][]{Yan2017,Cuneo2020}. Although the exact triggering mechanism of mini-outbursts and their state transitions still remain unclear, their properties of the observed quasi-periodic oscillations, state transitions, and light curves closely resemble those of main outbursts, suggesting that they may share the same physical processes \citep[e.g.,][]{Yan2017,Cuneo2020,Alabarta2022,Bright2025}.

MAXI J1348–630 is a recently discovered LMXB that has been high-cadence monitored across radio, optical, and X-ray bands \citep{2019ATelYatabe,2019ATelRussellopt,2019ATelRussellradio,2019ATelSanna,2019ATelChen}. MAXI J1348–630 underwent a main outburst beginning on January 27, 2019, during which it experienced a complete state transition and then returned to quiescence on May 15, 2019. Approximately 10 days later, it experienced a mini-outburst that remained in the hard state throughout, with a peak X-ray flux approximately one order of magnitude lower than the main outburst \citep{2019AtelNegoro,Jana2020,Tominaga2020,zhangliang2020,Carotenuto2021}. \citet{Weng2021} and \citet{You2024} have conducted comprehensive multi-wavelength studies of MAXI J1348-630 during its main outburst, exploring the radiative origins across different wavebands and the evolution of accretion flow. In this work, we present a multi-wavelength study on the mini-outburst of MAXI J1348–630 and aim to investigate the origin of its optical emission, which may help us further understand the state transition process in LMXBs. In Section \ref{sec:results}, we analyze the time delays between the optical and X-ray emissions of MAXI J1348–630, examine their flux correlation, and present the results of SED fitting. We discuss the implications for understanding the optical origin and state transitions in Section \ref{sec:discussion}.

\section{Data reduction}\label{sec:data}
% MAXI J1348-630 underwent a mini-outburst from about 2019-05-27 (MJD 58630) to 2019-08-05 (MJD 58700). The neutral hydrogen column density toward MAXI J1348-630 is taken to be $N_{\rm H}=8.6 \times 10^{21} \, {\rm cm}^{-2}$ \citep{Tominaga2020} and the distance of $2.2\, \rm kpc$ is adopted throughout this work \citep{Chauhan2021}.

\subsection{Insight-HXMT}
Insight-HXMT is China's first X-ray astronomical satellite lauched on 2017 June 15, carrying three main payloads with different effective areas \citep{HXMT-Zhang2020}: the Low Energy telescope (LE, $384  {\rm cm^2}$ at 1-12 keV), the Medium Energy telescope (ME, $952  {\rm cm^2}$ at 8-35 keV), and the High Energy telescope (HE, $5100 {\rm cm^2}$ at 20-350 keV). Insight-HXMT detected the mini-outburst of MAXI J1348-630 since MJD 58637 and has been continuously monitoring until MJD 58693. All X-ray data during the mini-outburst were reduced by Insight-HXMT Data Analysis Software \citep[HXMTDAS][]{HXMT-Zhang2020}. Following \citet{You2024}, the energy bands adopted in this work are 2-10 keV (LE), 10-28 keV (ME), and 28-150 keV (HE); data in 21–24 keV were ignored due to the photoelectric effect of electrons in the silver K shell. The light curves for LE, ME, and HE are shown in the top panel of Figure \ref{fig:lightcurve}.

During the mini-outburst of MAXI J1348-630, the Comptonization flux dominates the observed X-ray emission \citep{zhangliang2020}. Therefore, we performed the spectral fitting in the X-ray using the model {\tt tbabs*(diskbb+nthcomp)} in {\tt XSPEC} and then derived the X-ray Comptonization flux through {\tt cflux} module \citep{Arnaud_xspec}. The neutral hydrogen column density toward MAXI J1348-630 is taken to be $N_{\rm H}=8.6 \times 10^{21} \, {\rm cm}^{-2}$ \citep{Tominaga2020}. For X-ray spectral fitting of MAXI J1348–630 during its main outburst, we refer readers to \citet{You2024}. The light curve of X-ray Comptonization flux is presented at the bottom panel of Figure \ref{fig:lightcurve} and the data are tabulated in Table \ref{tab:xray_data}.

% $\textbf{F}_{\rm \textbf{comp}}$ \textbf{(erg}$\, \mathbf{s^{-1}\,cm^{-2}}$\textbf{)}
\begin{table}[htbp]
  \centering
  \caption{Comptonization flux of MAXI J1348-630}
  \begin{tabular}{*{2}{c}||*{2}{c}}
  \hline
  \textbf{MJD} & $\textbf{log F}_{\rm \textbf{comp}}$ & \textbf{MJD} & $\textbf{log F}_{\rm \textbf{comp}}$  \\
   & $(\rm erg\, s^{-1} \, cm^{-2})$ & & $(\rm erg\, s^{-1} \, cm^{-2})$ \\
  \hline
58637.39 & $-7.98\pm0.029$ & 58663.38 & $-7.59\pm0.008$ \\
58637.52 & $-7.93\pm0.016$ & 58663.38 & $-7.59\pm0.008$ \\
58637.65 & $-7.94\pm0.019$ & 58663.58 & $-7.58\pm0.010$ \\
58637.82 & $-7.97\pm0.015$ & 58663.58 & $-7.58\pm0.010$ \\
58639.08 & $-7.89\pm0.010$ & 58665.07 & $-7.62\pm0.009$ \\
58644.24 & $-7.57\pm0.008$ & 58665.21 & $-7.68\pm0.004$ \\
58648.48 & $-7.52\pm0.008$ & 58665.21 & $-7.68\pm0.004$ \\
58649.25 & $-7.50\pm0.007$ & 58665.36 & $-7.62\pm0.010$ \\
58649.38 & $-7.52\pm0.011$ & 58665.36 & $-7.62\pm0.010$ \\
58650.67 & $-7.48\pm0.009$ & 58667.06 & $-7.65\pm0.010$ \\
58650.84 & $-7.49\pm0.007$ & 58667.06 & $-7.65\pm0.010$ \\
58650.97 & $-7.60\pm0.007$ & 58667.22 & $-7.64\pm0.009$ \\
58651.86 & $-7.48\pm0.010$ & 58667.35 & $-7.65\pm0.011$ \\
58652.12 & $-7.57\pm0.026$ & 58670.06 & $-7.70\pm0.011$ \\
58652.13 & $-7.50\pm0.011$ & 58670.06 & $-7.70\pm0.011$ \\
58652.76 & $-7.46\pm0.008$ & 58670.20 & $-7.69\pm0.010$ \\
58652.82 & $-7.50\pm0.008$ & 58670.20 & $-7.69\pm0.010$ \\
58652.96 & $-7.52\pm0.015$ & 58672.28 & $-7.70\pm0.011$ \\
58653.88 & $-7.51\pm0.011$ & 58672.28 & $-7.70\pm0.011$ \\
58654.08 & $-7.50\pm0.012$ & 58672.45 & $-7.72\pm0.008$ \\
58654.11 & $-7.49\pm0.008$ & 58672.45 & $-7.72\pm0.008$ \\
58654.71 & $-7.49\pm0.008$ & 58672.61 & $-7.71\pm0.011$ \\
58654.81 & $-7.50\pm0.007$ & 58672.61 & $-7.71\pm0.011$ \\
58654.94 & $-7.49\pm0.010$ & 58672.74 & $-7.62\pm0.005$ \\
58655.37 & $-7.50\pm0.005$ & 58672.74 & $-7.62\pm0.005$ \\
58655.57 & $-7.48\pm0.009$ & 58673.94 & $-7.78\pm0.010$ \\
58655.70 & $-7.50\pm0.006$ & 58673.94 & $-7.78\pm0.010$ \\
58657.06 & $-7.52\pm0.008$ & 58675.33 & $-7.76\pm0.008$ \\
58657.06 & $-7.52\pm0.008$ & 58675.33 & $-7.76\pm0.008$ \\
58657.19 & $-7.52\pm0.007$ & 58676.92 & $-7.82\pm0.011$ \\
58657.33 & $-7.52\pm0.006$ & 58677.25 & $-7.86\pm0.011$ \\
58659.84 & $-7.56\pm0.016$ & 58679.82 & $-7.86\pm0.010$ \\
58659.84 & $-7.56\pm0.016$ & 58681.81 & $-7.88\pm0.013$ \\
58661.32 & $-7.57\pm0.007$ & 58681.95 & $-7.91\pm0.022$ \\
58661.32 & $-7.57\pm0.007$ & 58684.66 & $-7.96\pm0.012$ \\
58661.46 & $-7.55\pm0.008$ & 58687.17 & $-8.00\pm0.020$ \\
58661.46 & $-7.55\pm0.008$ & 58689.79 & $-8.19\pm0.019$ \\
58661.59 & $-7.57\pm0.008$ & 58691.41 & $-8.22\pm0.022$\\
58661.59 & $-7.57\pm0.008$ & 58691.54 & $-8.33\pm0.012$ \\
58663.22 & $-7.60\pm0.008$ & 58691.68 & $-8.33\pm0.034$ \\
58663.22 & $-7.60\pm0.008$ & 58693.40 & $-8.52\pm0.020$ \\
  \end{tabular}
\label{tab:xray_data}
\end{table}

\subsection{Las Cumbres Observatory}\label{subsec:LCO}
\begin{table}[htbp]
  \centering
  \caption{Observed optical magnitudes of MAXI J1348-630}
  \begin{tabular}{*{4}{c}}
  \hline
  \textbf{MJD} & $\mathbf{g'}$ & $\mathbf{r'}$ & $\mathbf{i'}$ \\
  \hline
    58624.64 & $18.96\pm0.14$ & $17.95\pm0.08$  & $17.12\pm0.11$   \\
    58630.44 & $17.83\pm0.09$ & $16.94\pm0.04$  & $16.30\pm0.04$   \\
    58631.08 & $17.76\pm0.10$ & $16.88\pm0.07$  & $16.25\pm0.07$   \\
    58634.77 & $17.42\pm0.10$ & $16.60\pm0.07$  & $15.94\pm0.07$   \\
    58639.42 & $17.13\pm0.06$ & $16.37\pm0.03$  & $15.67\pm0.06$   \\
    58640.01 & $17.21\pm0.08$ & $16.35\pm0.05$  & $15.72\pm0.07$   \\
    58640.48 & $16.99\pm0.02$ &  - &  $15.47\pm0.01$  \\
    58640.95 & $17.15\pm0.11$ & $16.29\pm0.07$ &  $15.66\pm0.10$  \\
    58641.96 & $17.07\pm0.08$ & $16.20\pm0.06$  & $15.54\pm0.05$   \\
    58643.70 & - & $16.38\pm0.10$  &  $15.78\pm0.07$  \\
    58643.95 & $17.21\pm0.09$ & $16.40\pm0.05$  & $15.75\pm0.08$   \\
    58645.36 & $17.25\pm0.03$ & $16.45\pm0.08$  & $15.78\pm0.02$   \\
    58647.71 & $17.28\pm0.14$ & $16.46\pm0.09$  & $15.81\pm0.08$   \\
    58648.37 & $17.27\pm0.10$ & $16.46\pm0.09$  & $15.82\pm0.04$   \\
    58649.82 & $17.26\pm0.14$ & $16.44\pm0.14$  & $15.75\pm0.13$   \\
    58651.40 & $17.26\pm0.09$ & $16.49\pm0.10$  & $15.85\pm0.06$   \\
    58652.82 & $17.36\pm0.12$ & $16.50\pm0.10$  & $15.82\pm0.09$   \\
    58655.36 & $17.37\pm0.10$ & $16.54\pm0.06$  & $15.87\pm0.06$   \\
    58656.97 & $17.37\pm0.08$ & $16.52\pm0.09$  & $15.87\pm0.08$   \\
    58662.00 & $17.46\pm0.09$ & $16.62\pm0.07$  & $15.96\pm0.07$   \\
    58664.96 & $17.48\pm0.08$ & $16.65\pm0.08$  & $16.00\pm0.08$   \\
    58673.72 & $17.62\pm0.10$ & $16.76\pm0.07$  & $16.09\pm0.09$   \\
    58675.74 & $17.58\pm0.12$ & $16.73\pm0.12$  & $16.05\pm0.1$   \\
    58680.75 & $17.59\pm0.19$ & $16.84\pm0.06$  & $16.23\pm0.09$   \\
    58686.83 & $17.99\pm0.10$ & $17.11\pm0.06$  & $16.49\pm0.07$   \\
    58693.80 & $18.44\pm0.12$ & $17.65\pm0.06$  & $16.99\pm0.07$   \\
    58697.80 & $18.73\pm0.12$ & $17.81\pm0.10$  & $17.05\pm0.12$   \\
    58698.80 & $18.82\pm0.16$ & $17.84\pm0.10$  & $17.06\pm0.09$   \\
    58701.80 & $19.11\pm0.12$ & $18.12\pm0.09$  & $17.38\pm0.08$   \\
    58704.75 & $19.51\pm0.26$ & $18.43\pm0.13$  & $17.62\pm0.12$   \\
  \end{tabular}
\label{tab:opt_data}
\end{table}

\begin{figure*}[htbp]
    \centering
    \includegraphics[scale=0.62]{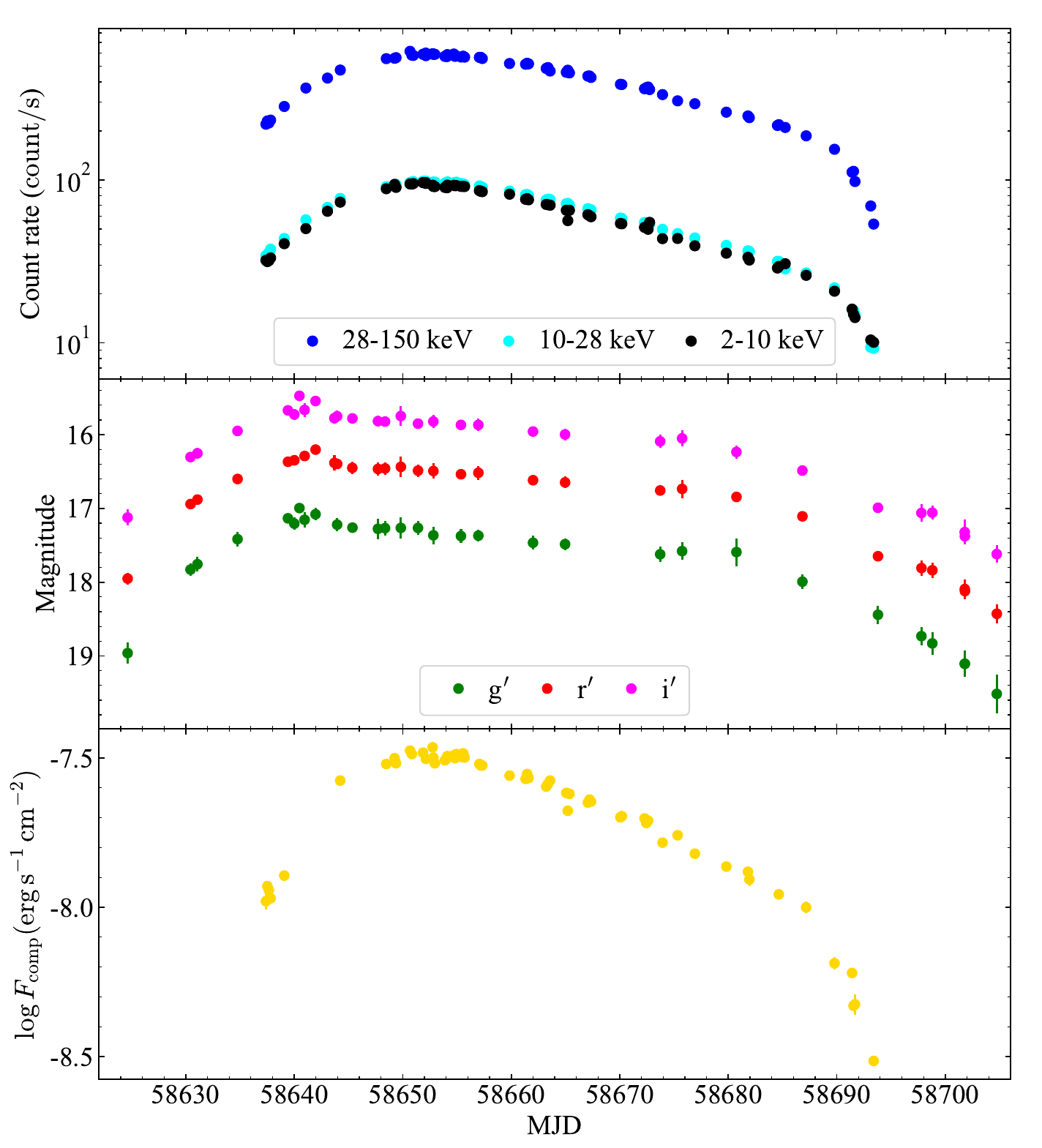}
    \caption{The light curves of MAXI J1348-630 during the mini-outburst. The top panel shows the X-ray count rates observed by Insight-HXMT and the middle panel presents the optical magnitudes observed by LCO. The bottom panel displays the Comptonization flux.}
    \label{fig:lightcurve}
\end{figure*}

The optical data were obtained from the Las Cumbres Observatory (LCO) Global Telescope, a worldwide network of telescopes designed for optical time-domain astronomy \citep{LCOGT}. LCO monitored the mini-outburst of MAXI J1348–630 starting from MJD 58624. Observations were carried out using the $g'$, $r'$, and $i'$ filters of the Sloan Digital Sky Survey (SDSS). The raw data were processed through the {\tt BANZAI} \footnote{\url{https://github.com/LCOGT/banzai?tab=readme-ov-file\#banzai-pipeline}} pipeline \citep{Banzai,Banzai2}, which provides standard image reduction. Photometric measurements were performed using the Automated Photometry of Transients pipeline \citep[AutoPhOT;][]{Autophot}, which calculates the AB magnitudes calibrated against the SkyMapper catalog and estimates the magnitude errors based on both the signal-to-noise ratio (SNR) and zero-point uncertainties. The observed $g'$, $r'$, and $i'$ band magnitudes are tabulated in Table \ref{tab:opt_data} and the corresponding light curves are shown in the middle panel of Figure \ref{fig:lightcurve}.

The optical magnitudes need to be corrected for the galactic extinction using the relations $A_{\rm g}=1.161 \times A_{\rm V}$, $A_{\rm r}=0.843 \times A_{\rm V}$, and $A_{\rm i}=0.639 \times A_{\rm V}$ \citep{1998ApJSchilegel}, where $A_{V}$ is derived from the hydrogen column density as $ A_{\rm V}=N_{\rm H}/2.21\times 10^{21} \, {\rm cm^{-2}}$ \citep{2009MNRASGuver} and $N_{\rm H}=8.6 \times 10^{21} \, {\rm cm}^{-2}$ \citep{Tominaga2020}. Only observations with $\mathrm{SNR} > 5$ were included in our subsequent analyses. 

\section{Results} \label{sec:results}
In this section, we investigate the time delays and flux correlation between the Comptonization flux and optical emission, and perform SED fitting for the X-ray and optical bands.
\begin{figure}
    \centering
    \includegraphics[scale=0.5]{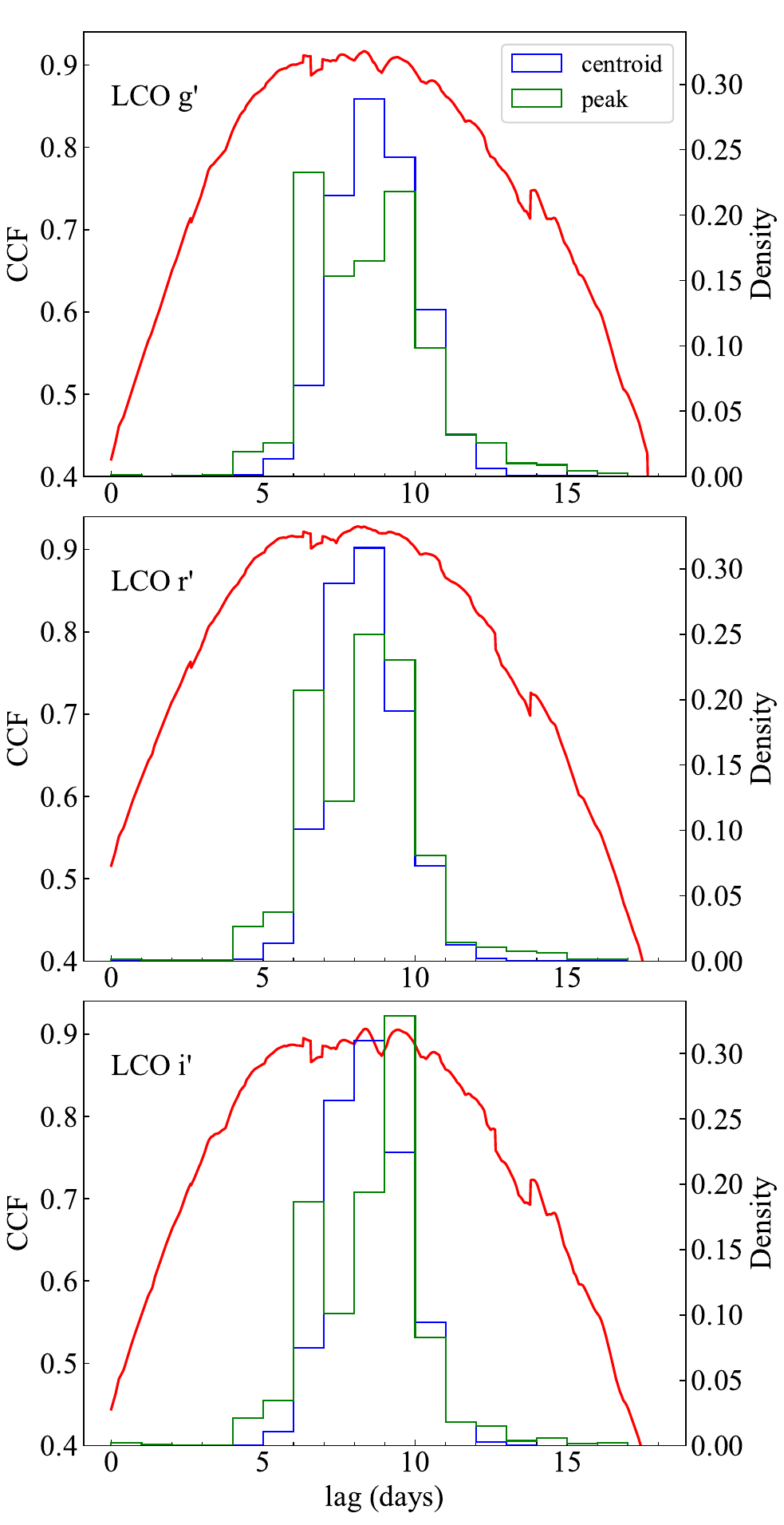}
    \caption{Cross-correlation analysis between the Comptonization flux and different optical bands ($g'$, $r'$, and $i'$ bands for the top, middle, and bottom panels, respectively). In all panels, the red line indicates the cross-correlation function (left axis), and the green and blue histograms are the cross-correlation peak delay distribution and the centroid delay distribution (right axis).}
    \label{fig:CCF}
\end{figure}

\subsection{Time delays}
Time delays between different radiation bands provide direct probes of the physical scales and underlying mechanisms in accreting systems. For example, whether the optical emission precedes the X-rays can be used to distinguish whether the optical radiation originates from X-ray reprocessing or synchrotron radiation of the hot accretion flow \citep{veledina2011,Veledina2017}. The delay between optical and X-ray emission may reflect the viscous timescale of the accretion disk, revealing how fluctuations propagate inward and outward \citep[e.g.,][]{Weng2021,You2023sience,DDZ2025}. 

As illustrated in Figure \ref{fig:lightcurve}, an obvious day-scale time delay can be found between the peaks of the optical and X-ray light curves. To quantitatively determine the time delay between optical emission and Comptonization flux, we employed an interpolated cross-correlation function (ICCF) analysis \citep{Gaskell1986,Gaskell1987}. ICCF allows us to estimate the time delay even for unevenly sampled data by interpolating one light curve to the epochs of the other. CCF/ICCF has been widely applied in reverberation mapping studies of AGNs \citep[e.g.,][]{Kaspi_RM_2000,Bentz_RM_2009,Du2014} and research on time delays across different wavebands in XRBs \citep[e.g.,][]{veledina2011,Veledina2017,You2023sience,You2024,DDZ2025}. For more detailed information on ICCF, readers are referred to \citet{Du2014} and \citet{You2023sience}. The uncertainties on time delays were calculated with the flux randomization/random subset sampling (FR/RSS) method \citep[described in][]{Peterson1998,Peterson2004}, which takes into account the measurement uncertainties and those arising from the sampling cadence. 

The ICCF results are presented in Figure \ref{fig:CCF}. We found that X-ray Comptonization light curve lags behind the $g'$ band by a peak delay of $\tau_{\rm p}=8.56^{+2.02}_{-2.05} \, {\rm days}$ days and the centroid delay of $\tau_{\rm c} =8.73^{+1.32}_{-1.32} \, {\rm days}$ (uncertainties are at the $1 \, \sigma$ level). For the $r'$ band, the delays are $\tau_{\rm p}=8.33^{+1.35}_{-1.82} \, {\rm days}$ and $\tau_{\rm c}=8.35^{+1.19}_{-1.15} \, {\rm days}$, while for the $i'$ band, they are $\tau_{\rm p}=8.54^{+1.32}_{-2.02} \, {\rm days}$ and $\tau_{\rm c}=8.54^{+1.22}_{-1.17} \, {\rm days}$. These results clearly indicate that the light curves of the three monochromatic optical bands are simultaneous, and that the X-ray emission lags behind the optical emission by approximately 8.5 days during the mini-outburst.

\subsection{Flux correlations}

The power-law correlations between the X-ray and optical flux ($F_{\rm opt} \propto F_{\rm X}^{\beta}$) can be used to probe the origin of optical emission in LMXBs \citep[e.g.,][]{Russell2006,Kosenkov2020}. A slope of $\beta \sim 0.7$ typically indicates synchrotron emission from a compact jet \citep{Gallo2003}; $\beta \sim 0.5$ implies that optical emission results from X-ray reprocessing in the outer accretion disk \citep{Paradijs1994}; and $\beta \sim 0.25$ suggests a viscously heated accretion disk origin of optical emission \citep[e.g.,][]{Yang2022}. Recent investigations of the correlations between X-ray Comptonization and optical flux in MAXI J1820+070 were conducted by \cite{You2023sience}, which found that $\beta \sim 1.0$ during the rising hard state, indicating a steeper correlation compared to those observed in the hard state of LMXBs, but no correlation was detected during the decaying hard state. These findings pose a challenge to our understanding of the origin of the optical emission.

\begin{figure}
    \centering
    \includegraphics[scale=0.42]{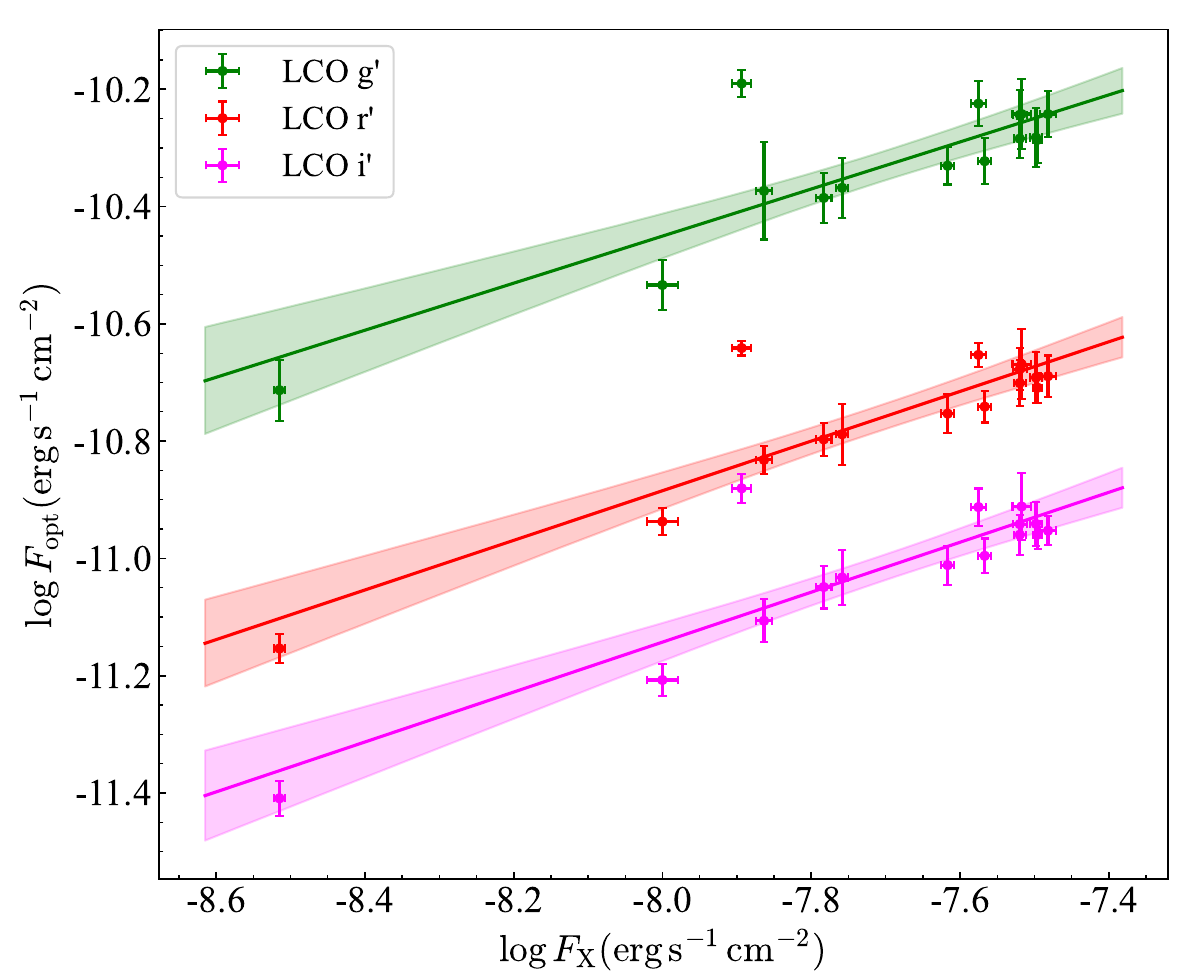}
    \caption{The power-law correlation between the Comptonization flux and different optical bands (see the legend). The solid lines are the best fits, and the shaded regions indicate uncertainties at the $1\, \sigma$ level.}
    \label{fig:corr}
\end{figure}

We selected quasi-simultaneous X-ray and optical observations (within one day) and fitted the correlation using the {\tt linmix} package, a Bayesian-based linear regression tool that takes flux errors into account \citep{linmix}. As shown in the top panel of Figure \ref{fig:corr}, a clear power-law correlation is observed between the Comptonization flux and the optical flux in mini-outburst of MAXI J1348-630, with best-fit slopes of $\beta=0.40 \pm 0.097$, $0.42 \pm 0.082$, and $0.43 \pm 0.083$ for $g'$, $r'$, and $i'$ bands, respectively. 
%\citet{You2024} suggested that it is necessary to consider the time delay in correlation analyses when delays are detected \citep[see also,][]{You2023sience}. Accordingly, we repeated the fits by taking into account the centroid time delays between the X-ray and optical bands, resulting in slopes of $\beta=0.41 \pm 0.088$, $0.35 \pm 0.064$, and $0.38 \pm 0.073$ for $g'$, $r'$, and $i'$ bands, respectively (see bottom panel of Figure \ref{fig:corr}). In both cases, with and without time delay correction, we consistently find $\beta \sim 0.4$ between the monochromatic optical and X-ray fluxes, which may be due to the limited X-ray data during the rising phase, resulting in most of the quasi-simultaneous data being concentrated in the decaying phase. Therefore, the effect of time delays may not significantly impact the correlation slopes.

% The top panel is based on quasi-simultaneous optical and X-ray observations (within one day), while the bottom panel takes into account the time delay correction between X-ray and optical bands.

\subsection{SED fitting}
We constructed SEDs using quasi-simultaneous optical and X-ray data (within one day) to investigate the multi-wavelength emission during the mini-outburst of MAXI J1348–630. In total, 15 quasi-simultaneous SEDs were collected. The SEDs were modeled using {\tt tbabs*diskir} in {\tt XSPEC}, where {\tt tbabs} accounts for X-ray absorption \citep{tbabs}, and {\tt diskir} represents an irradiated disk model, which considers the effects of inner disk and corona irradiation on the outer accretion disk \citep{diskir2008,diskir2009}. Note that {\tt tbabs} does not account for optical extinction; therefore, the optical data were corrected for extinction following the method described in Section~\ref{subsec:LCO}. The corrected optical data were then converted into {\tt XSPEC} format using {\tt ftflx2xsp} module \footnote{\url{https://heasarc.gsfc.nasa.gov/docs/software/lheasoft/help/ftflx2xsp.html}}. In the fittings, we fixed the neutral hydrogen column density toward MAXI J1348–630 at $N_{\rm H} = 8.6 \times 10^{21}, {\rm cm}^{-2}$ \citep{Tominaga2020}. There are nine parameters in the irradiated disk model; we fixed $f_{\rm in} = 0.1$ and $r_{\rm irr} = 1.1$ as the default value for the hard state \citep{diskir2008,diskir2009}. After testing, we found $L_{\rm c}/L_{\rm d}$ was poorly constrained, so we fixed it to the typical hard state value of 5 \citep[e.g.,][]{Rodi2021,Ozbey2022,Echiburu2024}. Finally, six parameters were free to fit the SEDs.

\begin{figure*}
    \centering
    \includegraphics[scale=0.54]{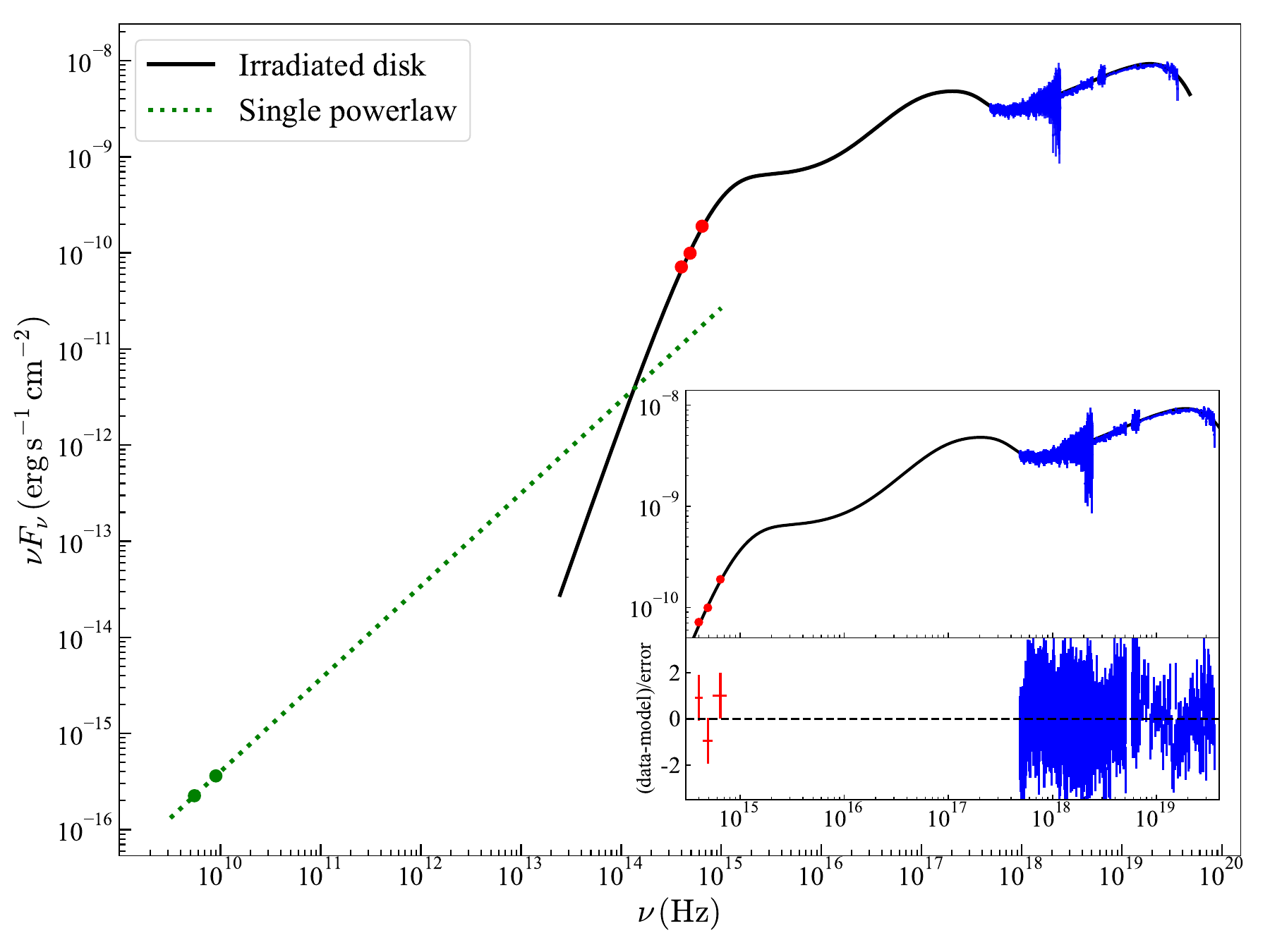}
    \caption{Best-fit unabsorbed SED of MAXI J1348-630 at MJD 58643. The blue and red data points correspond to the observation of Insight-HXMT and LCO. The radio data (green points) were observed by ACTA, presented by \citet{Carotenuto2021}. The black line represents the best-fit of model {\tt tbabs*diskir}. The green dotted line is a power-law fit to radio data. The bottom-right panel is a zoom-in of the X-ray and optical fits, along with the fit residuals.}
    \label{fig:SED}
\end{figure*}

\begin{figure*}
    \centering
    \includegraphics[scale=0.63]{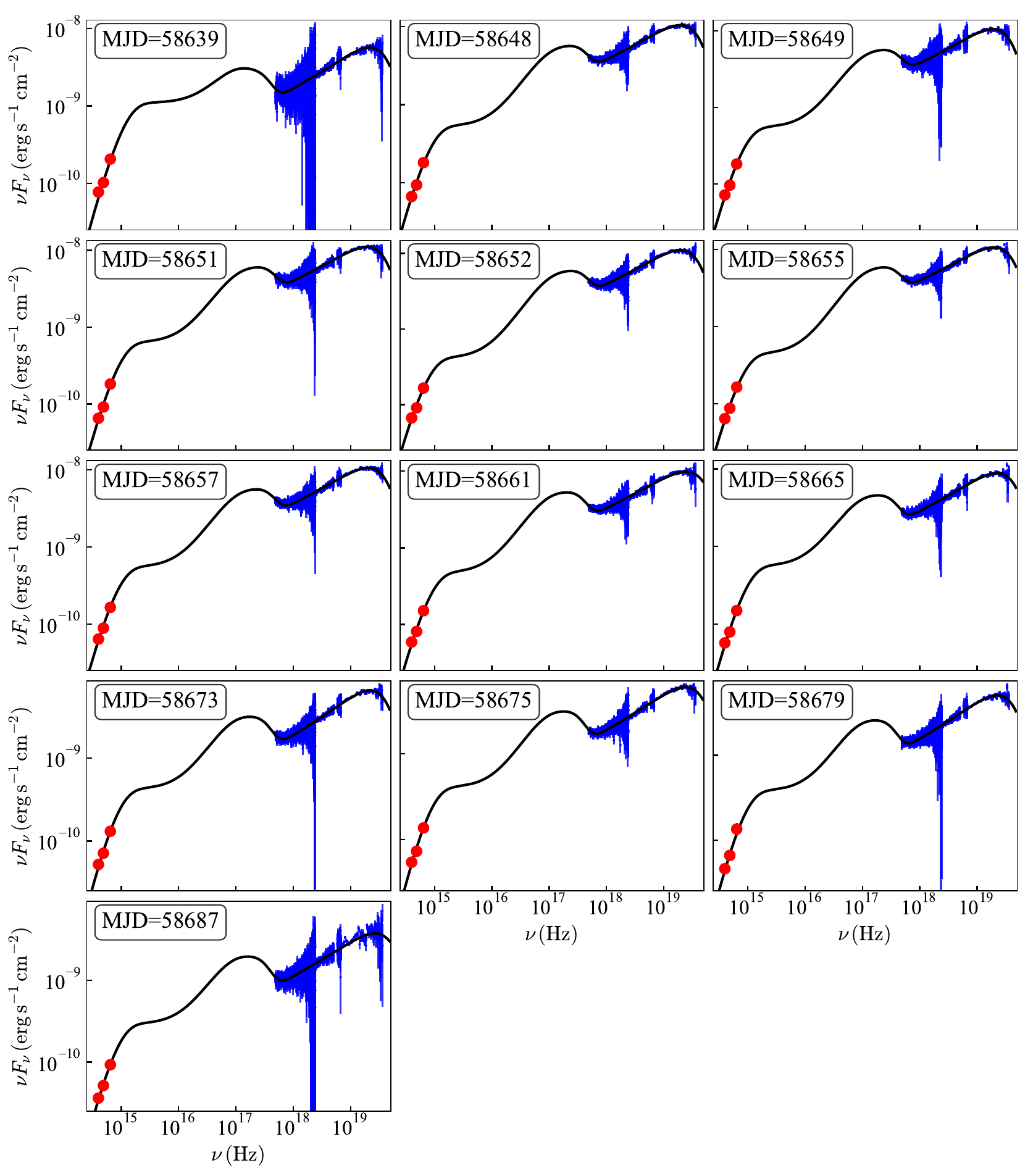}
    \caption{Best-fit unabsorbed SED of MAXI J1348-630. The blue and red data points correspond to the observation of Insight-HXMT and LCO. The black line represents the best-fit of model {\tt tbabs*diskir}.}
    \label{fig:SED_other}
\end{figure*}

Due to the poor quality of the X-ray spectrum on MJD 58693, we did not fit the SED for that day. We found that the remaining 14 quasi-simultaneous SEDs, excluding this one, can be well fitted by the irradiated disk model, and the best-fit parameters are reported in Table \ref{tab:fits}. As a representative case, in Figure \ref{fig:SED}, we displayed the SED and corresponding fitting results on MJD 58643, when happens to have quasi-simultaneous radio observations at 5.5 and 9 GHz \citep[][more about the radio emission is discussed in Section \ref{sec:origin of opt}]{Carotenuto2021}. The fitting results of the remaining quasi-simultaneous SEDs are shown in Figure \ref{fig:SED_other}. The well-fitting results clearly illustrate that the irradiated disk model can successfully account for the optical and X-ray emission of MAXI J1348–630 during its mini-outburst. The consistency of good fits throughout the mini-outburst further strengthens the interpretation that the optical emission is dominated by the irradiated outer disk.

\section{Discussions} \label{sec:discussion}
\subsection{Origin of Optical Emission} \label{sec:origin of opt}
%讨论时间延迟
Our results show that the optical emission significantly precedes the X-ray emission by $\sim 8.5$ days (see Figure \ref{fig:CCF}). Meanwhile, \citet{You2024} reported that there is no significant time delay between the X-ray and radio emission during the mini-outburst, implying that the optical emission should also lead radio emission by several days. The jet model predicts that the time delay between radio and optical emission is about tens of minutes \citep[e.g.,][]{Tetarenko2019}. Therefore, the analysis of the time delay disfavors the interpretation that the optical emission originates from the jet. On the other hand, if the optical radiation originates from X-ray reprocessing or synchrotron radiation of the hot accretion flow, the time delay between the optical and X-ray emissions is typically milliseconds to seconds \citep{veledina2011,Veledina2017}, which is inconsistent with our results. In fact, the X-ray peak lagging behind the optical peak by several days can be naturally interpreted by the DIM \citep{Lasota2001,2001Dubus,Goodwin2020,You2023sience,DDZ2025}. At the onset of an outburst, the region near the inner edge of the disk become ionized, accompanied by a local temperature rise; subsequently, two heating fronts propagate inward and outward \citep{Hameury2020}. When the heating front reaches the outer edge of the accretion disk, the optical emission soon peaks. However, the accretion flow continues to propagate inward due to viscous dissipation, causing the mass accretion rate at the truncation radius to reach its maximum at a later time \citep{DDZ2025}. Given that the Comptonization luminosity scales approximately with the square of the accretion rate at truncation radius, therefore, the X-ray peak is expected to lag behind the optical peak by approximately several days.

%讨论SED
\citet{Carotenuto2021} reported a flat radio spectrum of MAXI J1348-630 during the mini-outburst (i.e., $\alpha \sim 0$, where the radio flux density follows $L_{\nu} \propto \nu^{\alpha}$), which is the signature of self-absorbed synchrotron emission \citep{Blandford1979,Fender2001}, implying the presence of a compact radio jet. It has been suggested that the flat self-absorbed synchrotron spectrum of the compact jet could extend to near-infrared or even optical regimes \citep{Fender2001,Russell2013}, beyond which the jet becomes optically thin, exhibiting a spectral index of $-1 <\alpha<-0.5$. To estimate the contribution of the jet to the optical emission, we plotted the quasi-simultaneous radio data of 5.5 and 9 GHz presented by \citet{Carotenuto2021} in Figure \ref{fig:SED} (green dots). The frequency at which self-absorbed synchrotron radiation transitions from optically thick to optically thin remains highly uncertain \citep[e.g.,][]{Russell2013}. Therefore, for simplicity, we fit the radio data with a single power-law ($\alpha =-0.03$, corresponding to a spectral index of 0.97 in Figure \ref{fig:SED}) and extrapolate it to the optical band (green dotted line). Given that the synchrotron radiation of the jet is possibly optically thin in the optical band, the contribution of the jet to optical emission must be lower than the green dotted line. As shown in the figure, this line still lies significantly below the optical data, suggesting that the contribution of the jet to the optical emission during the mini-outburst of MAXI J1348–630 is negligible. 

As shown in Figure \ref{fig:SED} and \ref{fig:SED_other}, there is a significant excess of optical emission above the standard disk spectrum. \citet{Veledina2013} suggested that this excess could originate from synchrotron radiation produced by non-thermal electrons in the extended hot accretion flow. This model predicts that the optical/infrared (OIR) spectrum should be flat, with a spectral index $\alpha_{\rm OIR}$ ranging from approximately -0.5 to 0.5 (where $L_{\nu} \propto \nu^{\alpha_{\rm OIR}}$), which arises from the superposition of synchrotron self-absorption peaks from different radial zones of the hot flow. However, our optical data reveal a spectral slope of $\alpha_{\rm OIR} \sim 1$, which deviates from the flat spectrum expected from the hot flow synchrotron radiation scenario. This discrepancy suggests that the optical emission is unlikely to be dominated by synchrotron radiaition from the hot accretion flow.

Our SED fitting results indicate that the optical emission can be well fitted by the irradiation model. However, if irradiation were the dominant mechanism, there should be no discernible time delay between the X-ray and optical light curves. It should be noted that irradiation is also incorporated in the DIM \citep[e.g.,][]{You2023sience,DDZ2025}. X-ray irradiation elevates the disk temperature, causing the disk SED to deviate from the steady-state multi-color blackbody solution (where $\nu F_{\nu} \propto \nu^{4/3}$). The temperature increase enhances the gas pressure in the disk, ultimately leading to an elevated accretion rate and subsequent additional energy release. In the DIM simulation, as demonstrated in Figure 4 of \citet{DDZ2025}, the effective temperature of the accretion disk surpasses the temperature contribution from irradiation. Therefore, the viscosity enhanced by irradiation serves as the primary source of the optical radiation. Under these conditions, the SED of the accretion disk no longer follows the steady-state disk solution but rather resembles the SED under irradiation \citep[see Figure 8 of][]{DDZ2025}. For the sake of simplicity, this work employs the irradiation disk model to fit the SED as an approximation under non-steady-state. Therefore, during the mini-outburst, we can only conclude that the optical radiation originates from disk rather than jet or hot accretion flow, without being able to further specify whether it comes from viscous heating or X-ray reprocessing.

\begin{table*}[htbp]
  \centering  
  \begin{threeparttable}
  \caption{Best-fit values of parameters and errors at 90\% confidence level with model {\tt tbabs*diskir}}
  \label{tab:fits}
  \begin{tabular}{*{8}{c}} 
  \hline
  \addlinespace[0.3em]
    \textbf{MJD} & $\mathbf{kT_{\rm \textbf{in}}}$ \textbf{(keV)} & $\mathbf{\Gamma}$ & $\mathbf{kT_{\rm \textbf{e}}}$ \textbf{(keV)} & $\mathbf{f_{\rm \textbf{out}}}$ \textbf{(}$\mathbf{10^{-3}}\textbf{)}$ & \textbf{log(}$\mathbf{r}_{\rm \textbf{out}}$\textbf{/}$\mathbf{r}_{\rm \textbf{in}}$\textbf{)} & \textbf{Norm} \textbf{(}$\mathbf{10^{5}}\textbf{)}$ & $\mathbf{\chi^2}$\textbf{/d.o.f} \\
    \addlinespace[0.3em]
  \hline
    \addlinespace[0.3em]
    58639 & $0.179^{+0.009}_{-0.009}$ & $1.593^{+0.016}_{-0.015}$ & $57.13^{+10.23}_{-6.88}$ & $23.4^{+23.9}_{-10.1}$ & $3.85^{+0.12}_{-0.13}$ & $1.56^{+0.38}_{-0.31}$ & 1547/1277 \\
    \addlinespace[0.3em]
    58643 & $0.179^{+0.002}_{-0.003}$ & $1.632^{+0.008}_{-0.008}$ & $49.94^{+2.95}_{-2.48}$ & $8.2^{+9.9}_{-3.7}$ & $3.81^{+0.14}_{-0.14}$ & $2.74^{+0.21}_{-0.14}$ & 1417/1277 \\
    \addlinespace[0.3em]
    58648 & $0.181^{+0.003}_{-0.002}$ & $1.671^{+0.007}_{-0.007}$ & $58.89^{+3.82}_{-3.43}$ & $6.9^{+3.9}_{-2.5}$ & $3.79^{+0.09}_{-0.11}$ & $2.97^{+0.14}_{-0.19}$ & 1473/1277 \\
    \addlinespace[0.3em]
    58649 & $0.188^{+0.004}_{-0.005}$ & $1.663^{+0.013}_{-0.013}$ & $54.35^{+5.99}_{-4.81}$ & $6.7^{+11.5}_{-3.7}$ & $3.84^{+0.22}_{-0.24}$ & $2.56^{+0.27}_{-0.23}$ & 1308/1277 \\
    \addlinespace[0.3em]
    58651 & $0.188^{+0.004}_{-0.004}$ & $1.655^{+0.011}_{-0.011}$ & $52.64^{+4.72}_{-3.77}$ & $7.4^{+9.1}_{-3.4}$ & $3.76^{+0.13}_{-0.14}$ & $2.81^{+0.28}_{-0.23}$ & 1379/1277 \\
    \addlinespace[0.3em]
    58652 & $0.186^{+0.003}_{-0.003}$ & $1.664^{+0.009}_{-0.009}$ & $56.60^{+4.78}_{-3.94}$ & $5.1^{+6.8}_{-2.4}$ & $3.84^{+0.17}_{-0.17}$ & $2.82^{+0.22}_{-0.19}$ & 1433/1277 \\
    \addlinespace[0.3em]
    58655 & $0.185^{+0.003}_{-0.003}$ & $1.665^{+0.008}_{-0.008}$ & $58.51^{+4.35}_{-3.55}$ & $4.9^{+4.8}_{-1.9}$ & $3.82^{+0.13}_{-0.13}$ & $2.86^{+0.18}_{-0.16}$ & 1390/1277 \\
    \addlinespace[0.3em]
    58657 & $0.183^{+0.004}_{-0.003}$ & $1.658^{+0.009}_{-0.010}$ & $60.60^{+5.53}_{-4.66}$ & $6.5^{+7.5}_{-2.7}$ & $3.78^{+0.13}_{-0.15}$ & $2.92^{+0.24}_{-0.23}$ & 1367/1277 \\
    \addlinespace[0.3em]
    58661 & $0.171^{+0.003}_{-0.003}$ & $1.671^{+0.010}_{-0.010}$ & $90.56^{+20.97}_{-13.36}$ & $5.7^{+7.6}_{-2.4}$ & $3.73^{+0.13}_{-0.16}$ & $3.52^{+0.32}_{-0.28}$ & 1360/1277 \\
    \addlinespace[0.3em]
    58665 & $0.164^{+0.003}_{-0.003}$ & $1.652^{+0.010}_{-0.010}$ & $74.26^{+11.25}_{-8.46}$ & $7.8^{+10.3}_{-3.3}$ & $3.68^{+0.13}_{-0.15}$ & $3.67^{+0.34}_{-0.30}$ & 1280/1277 \\
    \addlinespace[0.3em]
    58673 & $0.162^{+0.004}_{-0.004}$ & $1.631^{+0.010}_{-0.010}$ & $68.74^{+8.24}_{-6.40}$ & $8.5^{+13.0}_{-4.0}$ & $3.76^{+0.16}_{-0.17}$ & $2.65^{+0.28}_{-0.23}$ & 1365/1277 \\
    \addlinespace[0.3em]
    58675 & $0.163^{+0.003}_{-0.003}$ & $1.615^{+0.009}_{-0.009}$ & $61.96^{+6.06}_{-5.00}$ & $8.3^{+15.0}_{-4.3}$ & $3.78^{+0.19}_{-0.19}$ & $2.63^{+0.20}_{-0.18}$ & 1214/1277 \\
    \addlinespace[0.3em]
    58679 & $0.158^{+0.004}_{-0.004}$ & $1.635^{+0.010}_{-0.010}$ & $77.82^{+13.49}_{-9.50}$ & $9.3^{+16.5}_{-5.5}$ & $3.75^{+0.23}_{-0.23}$ & $2.45^{+0.28}_{-0.23}$ & 1379/1277 \\
    \addlinespace[0.3em]
    58687 & $0.154^{+0.007}_{-0.007}$ & $1.641^{+0.021}_{-0.021}$ & $112.96^{+162.68}_{-38.06}$ & $9.5^{+13.0}_{-4.2}$ & $3.75^{+0.15}_{-0.16}$ & $2.02^{+0.53}_{-0.47}$ & 1341/1277 \\
    \addlinespace[0.3em]
  \end{tabular}
  \begin{tablenotes}
    \small
    \item \textbf{Notes.} 
    $kT_{\rm in}$: innermost temperature of the disk (the same as in {\tt diskbb}); 
    $\Gamma$: power-law photon index; 
    $kT_{\rm e}$: electron temperature; 
    $f_{\rm out}$: the irradiation fraction of the bolometric flux;
    $r_{\rm out}$: the outer radius of the disk ; 
    Norm: disk component normalization (the same as in {\tt diskbb}); 
  \end{tablenotes}
  \end{threeparttable}
\end{table*}

%讨论correlation
The slope of the correlation between optical and X-ray emission, $\beta \sim 0.4$, is significantly lower than the predicted value of 0.7 based on the jet origin model \citep{Gallo2003}. For the model that optical emission originates from synchrotron radiation of the hot accretion flow, a relatively complex optical-X-ray correlation is predicted, which depends on the electron distribution and the slope of SED \citep{Kosenkov2020}. For the typical parameters of a radiatively inefficient accretion flow, the value is $\beta \sim 0.78$, which also shows a significant discrepancy from our results. Our results lie between the predicted value of $\sim 0.25$ based on viscous heating and $\sim 0.5$ predicted by X-ray reprocessing \citep[see][and the references therein]{Russell2006}, suggesting that $\beta \sim 0.4$ could be achieved when both irradiation and viscous heating contribute to the optical emission, which is consistent with DIM.

Note that the correlation slope we obtained is quite similar to that observed in the mini-outburst of MAXI J1820+070 \citep{Bright2025,Yang2025} and in GX 339-4 during its rising hard state \citep{Kosenkov2020}. However, there are significant differences in the results of their SED fitting: the optical emission of the MAXI J1820+070 is attributed to the jet \citep{Ozbey2022}, whereas that of GX 339-4 originates from synchrotron radiation of the hot accretion flow \citep{Kosenkov2020}. In contrast, our SED fitting results suggest that the optical emission in the mini-outburst of  MAXI J1348-630 is more likely come from the disk. Therefore, we argue that relying solely on flux correlation analysis may be insufficient to conclusively determine the origin of optical emission in LMXBs.

In summary, our analysis favors the interpretation that the optical emission of MAXI J1348-630 during the mini-outburst originates from the disk rather than the jet or hot accretion flow, and DIM is the key to understanding the optical emission that precedes the X-rays and the observed SED.

\subsection{Color-magnitude diagram}
\begin{figure}
    \centering
    \includegraphics[scale=0.38]{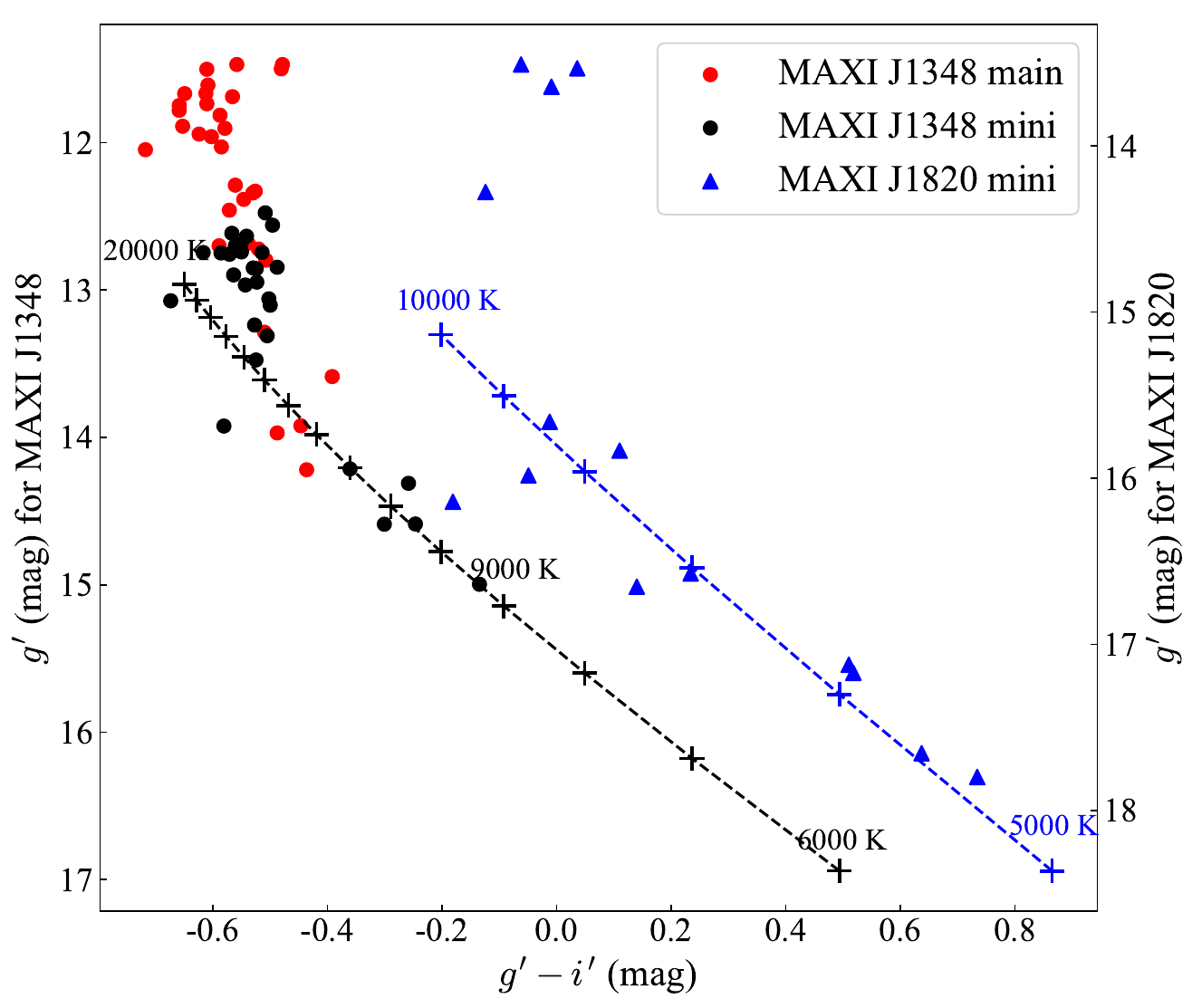}
    \caption{Color–magnitude diagram of MAXI J1348–630 and MAXI J1820+070. The magnitudes in this figure have been corrected for extinction. The left and right y-axes represent the $g^\prime$ magnitudes of MAXI J1348–630 and MAXI J1820+070, respectively. Red and black points denote the main outburst and mini-outburst of MAXI J1348–630, while blue triangles represent the mini-outburst of MAXI J1820+070. The two dashed lines are the theoretical curves of a single-temperature blackbody for each source \citep[see][for more details]{Maitra2008}, and the plus signs mark the temperature of the blackbody at intervals of $1000 \, \rm{K}$.}
    \label{fig:CMD}
\end{figure}

To investigate the color evolution of MAXI J1348–630 during its mini-outburst, we present the color–magnitude diagram \citep[CMD,][]{Maitra2008} based on the magnitudes of the $g'$ and $i'$ bands. CMD of MAXI J1348–630 are shown in Figure \ref{fig:CMD}. It can be seen that at lower luminosities, the evolution of MAXI J1348–630 roughly follows the expectation of a single-temperature blackbody with increasing temperature \citep[for more details about the methodology of CMD, see][]{Maitra2008}, but as the luminosity increases, it gradually begins to deviate. This trend is quite similar to that observed in some other LMXBs \citep[e.g.,][]{Russell2011,Poutanen2014,Zhang2019}, and we suggest that this deviation may be due to the fact that the single-temperature blackbody model oversimplifies the temperature profile of the accretion disk. However, MAXI J1348–630 never exhibits a prominent ``hook" feature \citep[see figure in][]{Russell2011,Poutanen2014}, which has been interpreted as a signature of jet contribution. This further supports our argument that the optical emission of MAXI J1348–630 is dominated by the disk rather than by the jet. \citet{Ozbey2022}, when fitting the SED of MAXI J1820+070 during its mini-outburst using an irradiated disk model, found that a good fit could only be achieved by including an additional power-law component extending from the radio to the optical band, which was interpreted as a contribution of the jet. For comparison, we also plotted the color evolution of the first mini-outburst of MAXI J1820+070 (MJD from 58560 to 58670), where optical data are taken from \citet{Sai2021}. The trend of MAXI J1820+070 is similar to that of MAXI J1348–630, but its color index is redder overall, which may be due to a lower disk temperature or the contribution of the jet.

We also analyzed the optical data of MAXI J1348–630 during its main outburst and present its color evolution in Figure \ref{fig:CMD} for comparison. Unlike some other LMXBs \citep[e.g.,][]{Russell2011,Zhang2019,2023Saikia}, the evolutionary trend of MAXI J1348–630 during the main outburst does not follow the expectation of a single-temperature blackbody; however, this trend appears to be consistent with the high-luminosity portion of the mini-outburst. On the other hand, \citet{Weng2021} suggested that the optical emission of MAXI J1348–630 during the main outburst originates from the accretion disk, and reported an optical and X-ray correlation slope of $\sim 0.38$. This value is very similar to that found in the mini-outburst, which may imply that the optical origins during the main outburst and the mini-outburst are possibly the same. In fact, it has been suggested that the mini-outburst and the main outburst are driven by the same mechanism \citep[e.g.,][]{Yan2017,Cuneo2020}, but more research is required to investigate the relationship between the mini-outburst and the main outburst.

\subsection{On the state transition}
As observed in most mini-outbursts of LMXBs \citep[e.g.,]{Patruno2016,Zhang2019,Ozbey2022,2023Saikia}, MAXI J1348–630 remained in the LHS throughout its mini-outburst without undergoing a state transition \citep{zhangliang2020,Carotenuto2021}. The truncated disk model serves as the paradigm for understanding state transitions \citep{ARAADone2007,ARAAYuan2014}, in which a geometrically thin, optically thick accretion disk is truncated at a certain radius. Inside the truncation radius, the accretion disk transitions to a hot, optically thin, geometrically thick accretion flow, often modeled as an advection-dominated accretion flow (ADAF) or a corona. As the source transitions from the LHS to the HSS, the truncation radius gradually moves inward toward the innermost stable circular orbit (ISCO), significantly increasing its thermal disk emission. Concurrently, the hot accretion flow (or corona) diminishes in size and luminosity. The reverse processes occur during the transition back to the LHS, where the disk starts to recede, reducing the thermal disk emission, and the Comptonized power-law component once again dominates the X-ray spectrum. This process for the soft-to-hard state transition was recently confirmed with the observations of MAXI J1820+070 \citep{You2023sience}. \citet{Dai2020} found that the disk of MAXI J1348-630 actually moved inward during the mini-outburst, but never reached the ISCO, consequently failing to transition to the soft state \citep[see also,][]{Garcia2019}.

The mini-outburst of MAXI J1348–630 resembles a ``failed outburst" \citep{zhangliang2020}. Recent studies have revealed that failed outbursts, during the several tens of days after the onset of the outburst, exhibit X-ray light curves, HID tracks, and timing properties that are nearly indistinguishable from those of full outbursts, suggesting a common physical mechanism at the onset \citep{Kosenkov2020,Alabarta2021}. However, \citet{Lucchini2023} suggested that X-ray variability, especially changes in power spectral hue, tends to evolve 10 to 40 days prior to a state transition. In contrast, such an evolution is not observed in failed outbursts. This finding could serve as a predictive tool for anticipating state transitions and may provide new insights into the physical mechanisms of state transitions. On the other hand, in the optical and infrared bands, \citet{Alabarta2021} found that failed outbursts are brighter than full outbursts, and \citet{Yang2025} showed that optical and X-ray light curves dip before the hard-to-soft transition. Therefore, optical emission may be the key to understanding state transitions.

Although state transitions of mini-outbursts have been reported in several LMXBs \citep[e.g., GRS 1739-278 and MAXI J1535-571,][]{Yan2017,Cuneo2020}, the lack of optical observations precludes a direct comparison with MAXI J1348-630 to further investigate state transitions. In fact, optical emission of many mini-outbursts has been missed due to their low luminosity and/or the lack of regular optical monitoring. We look forward to more sustained, high-cadence, multi-wavelength monitoring of LMXBs, which will enable further investigation of the state transition of LMXBs.

\begin{acknowledgments}
\section*{Acknowledgments}
This work is supported by Natural Science Foundation of China (NSFC) grants 12322307, 12361131579, and 12273026; by Xiaomi Foundation / Xiaomi Young Talents Program.
\end{acknowledgments}

\software{XSPEC \citep{Arnaud_xspec}, linmix \citep{linmix}, AutoPhoT \citep{Autophot}, BANZAI pipeline \citep{Banzai,Banzai2}, HXMTDAS \citep{HXMT-Zhang2020}}

\bibliography{1348}{}
\bibliographystyle{aasjournal}

\end{document}